\documentclass[useAMS,usenatbib]{mn2e}
\usepackage{psfig}

\def\ledd{L_{\rm Edd}}
\def\medd{\dot{M}_{\rm Edd}}

\newcommand\mdot{\dot{m}}
\newcommand\mbh{M_{\rm BH}}
\newcommand\md{M_{\rm d}}
\newcommand\msun{{\,M_\odot}}
\newcommand\lsun{{\,L_\odot}}
\newcommand\rsun{{\,R_\odot}}
\newcommand\sch{Schwarzschild~}
\newcommand\fe{Fe K$\alpha$\ }

\newcommand\teff{T_{\rm eff}}

\newcommand\rsg{R_{\rm sg}}

\newcommand\rout{R_{\rm out}}

\newcommand\sgra{Sgr~A$^*$}
\newcommand\schodel{Sch\"odel~}

\def\>{$>$}
\def\<{$<$}

\def\simlt{\lower.5ex\hbox{$\; \buildrel < \over \sim \;$}}
\def\simgt{\lower.5ex\hbox{$\; \buildrel > \over \sim \;$}}
\def\sqr#1#2{{\vcenter{\hrule height.#2pt
      \hbox{\vrule width.#2pt height#1pt \kern#1pt
         \vrule width.#2pt}
      \hrule height.#2pt}}}

\def\del#1{{}}

\title[Close stars in LLAGN]{Close stars and accretion in Low
Luminosity Active Galactic Nuclei}

\author[Nayakshin]{Sergei Nayakshin \\ Max-Planck-Institut f\"ur
Astrophysik, Karl-Schwarzschild-Str. 1, 85740 Garching, Germany}
\date{2004 Xxxxx XX}

\pagerange{\pageref{firstpage}--\pageref{lastpage}} \pubyear{2003}

\def\LaTeX{L\kern-.36em\raise.3ex\hbox{a}\kern-.15em
    T\kern-.1667em\lower.7ex\hbox{E}\kern-.125emX}

\begin{document}

\label{firstpage}

\maketitle

\begin{abstract}
Quasar accretion disks are believed to form stars by self-gravity.
Low Luminosity Active Galactic Nuclei (LLAGN) are much dimmer galactic
centers, and are often believed to be quasars that ran out of gaseous
fuel. LLAGN accretion disks should thus co-exist with thousands to
millions of stars or proto-stars left from the previous stronger
accretion activity. In principle, these stars may produce several
important effects: (i) contribute to the optical/UV spectra of some
LLAGN; (ii) reprocessing of the stellar radiation in the dusty disks
could dominate the LLAGN infra-red spectra; (iii) deplete the
(accretion) gas disk much faster than it can accrete onto the
supper-massive black hole (SMBH); (iv) stars, individually or in
groups, may slow down and modulate the accretion flow significantly
due to their inertia. In this way they may produce the LLAGN cut-off
disks; (v) alternatively, frequent enough stellar collisions and
resulting stellar disruptions could keep the inner disk empty. Here we
explore these ideas. We find that, despite ``low'' luminosities of
LLAGN, unrealistically high stellar densities are required to make a
sizable radiative contribution to the (HST) optical/UV spectra of
these galactic nuclei. Stellar contribution to the infrared spectrum
is more likely. Further, if LLAGN are in a quasi steady-state for as
long as $10^7$ years or more, too high stellar densities would again
be required to significantly affect the dynamics of accretion
flow. However, if LLAGN are ``short''-lived phenomena, e.g. $t \simlt
10^5$ years, quiescent states of quiescence-outburst cycles, then
embedded stars may be much more important through the mass effects
(iii) -- (v). With observations of LLAGN becoming progressively
better, it will be more and more difficult to neglect the presence of
close stars in and around nuclear accretion disks.
\end{abstract}

\section{Introduction}

LLAGN are galactic centers intermediate in luminosity and in some key
properties between quasars (bolometric luminosity $L_{\rm b}\sim
10^{46}$ erg s$^{-1}$) and inactive galactic centers (IGC), such as
our own (\sgra), whose luminosity is about $L_{\rm b}\sim 10^{36}$ erg
s$^{-1}$.  LLAGN are also known as LINERs (low ionization nuclear
emission-line regions; Heckman 1980) or as ``dwarf'' Seyfert nuclei
(e.g. Ho, Filippenko \& Sargent 1997, 2003). Typical nuclear
bolometric luminosities of these sources are $L_b \simlt 10^{42}$ erg
s$^{-1}$ (e.g. Ho 2003).  It is well known that most present day
galaxies harbor supper massive black holes (SMBHs), implying that
LLAGN and IGC too had to be bright accreting sources at some earlier
times. LLAGN and IGC are thus quasar relics or quiescent phases of a
long-term evolution of galactic nuclei. Some of the processes
operating in the quasar phase may have long-lasting effects and
continue to influence SMBH activity in the present dim epoch.  One
such process is star formation inside quasar accretion disks due to
self-gravity at distances greater than $\sim 10^3$ \sch radii
(e.g. Paczy\'nski 1978; Kolykhalov \& Sunyaev 1980, Shlosman \&
Begelman 1989; Goodman 2003), where $R_S = 2 G M_{\rm BH}/c^2$, the
\sch radius for the black hole with mass $M_{\rm BH}$. The goal of
this paper is to investigate some of the possible effects of the
presence of these close and fast moving stars on the LLAGN
phenomenon.

Let us summarize the observational and theoretical framework in which
bright AGN, LLAGN and inactive galactic centers are currently
understood.  Quasar accretion disks are believed to be the
``standard'' accretion disks (Shakura \& Sunyaev 1973), possibly
supplemented by an X-ray emitting corona. The cold accretion disk
apparently extends down to the last stable orbit, as evidenced by a
very prominent thermal-like UV feature called the Big Blue Bump (BBB)
and by the broad \fe lines observed in many cases (it is our personal
opinion that the non-detection of broad \fe lines in other sources is
caused by an over-ionization of the accretion disk surface). The
near-Eddington luminosities of quasars, short time scale variability,
and the high radiative efficiency with which most of the SMBHs are
inferred to have grown (Yu \& Tremaine 2002) are also best explained
in the context of the standard accretion disks since they indeed have
very high radiative efficiency.

LLAGN, and our Galactic Center (GC) black hole, \sgra, are distinctly
different in this regard, possibly requiring a radiatively inefficient
accretion. Chandra X-ray observations constrain the amount of hot gas
at the capture radius in \sgra\ and M87 (Baganoff et al. 2003; Di
Matteo et al. 2003), which allows one to estimate the Bondi accretion
rate onto the SMBHs. For other sources where direct X-ray data are not
constraining enough, Ho (2003) notes that the centers of the galaxies
hosting LLAGN should be filled with hot winds from young stars
(similar to our GC), yielding an accretion rate estimate. In addition,
the accretion rate through a cold accretion disk may be estimated by
using the optical/UV data in some LLAGN (e.g. Quataert et al. 1999;
Ptak et al. 2004). All of these methods lead to the same conclusion --
the expected luminosity from a thin standard accretion disk fed at the
deduced accretion rate is some $1-4$ orders of magnitude higher than
actually observed.

In addition, in stark contrast to quasars and Seyfert galaxies, LLAGN
do not show the BBB in their spectra (Ho 1999), have no short time
scale X-ray variability and no broad \fe lines (e.g. Dewangan et
al. 2004; Ptak et al. 2004). This strongly argues for the absence of
the cold accretion disks in the inner $\sim 10^2-10^3
R_S$. Theoretical interpretation of this fact is that the cold disk
``evaporates'' at few hundred to few thousand $R_S$ (e.g. Quataert et
al. 1999; Fig. 2 in Ho 2003; Ptak et al. 2004) to form an inner hot
Non-Radiative Accretion Flow (e.g. Narayan \& Yi 1994; Blandford \&
Begelman 1999). The most frequently discussed evaporation mechanism is
that of Meyer \& Meyer-Hofmeister (1994). If the ions of the hot flow
are much hotter than the electrons, then the flow is radiatively
inefficient and is much less luminous than the standard flow,
explaining the discrepancy. In the least luminous (in terms of $L_{\rm
b}/\ledd$, where $\ledd = 1.3 \times 10^{46} M_8$ erg s$^{-1}$, the
Eddington limit for the black hole mass $M_8 = \mbh/10^8 \msun$)
cases, such as \sgra, it is possible that a cold accretion disk does
not even exists (e.g. Falcke \& Melia 1997; Narayan 2002).

The best studied example of an inactive galactic center is the center
of our own galaxy, \sgra. Genzel et al. (2003; see also Levin \&
Beloborodov 2003) discovered that most of hot young ($t < 10^7$ years)
stars in the GC lie in one of the two {\em stellar disks}, strongly
suggesting these stars were formed by the self-gravity in two massive
accretion disks existing in earlier times. In the case of the GC,
these young stars actually dominate and determine the SMBH activity,
since their radiative output is some $\sim 5$ orders of magnitude
greater than that of the accretion flow, and since the hot winds
produced by these stars are believed to be the main source of fuel for
\sgra\ (e.g.  Coker \& Melia 1997).

It is thus likely that important interactions between close stars and
the accretion flow may be taking place in LLAGN as well. Star
formation may even continue in the LLAGN phase as well (e.g., see
Fig. 1 in Goodman 2003; also Milosavljevic \& Loeb 2004). Quite
generally, close stars should assume more prominent roles in the LLAGN
(and IGC) phase than they do in quasars because the mass and the
luminosity output of the gaseous accretion disk are modest in these
sources. The possible effects of the close stars can be divided into
two categories: the luminosity and the mass effects. The former are
due to the combined stellar luminosities; whereas the latter are due
to stellar mass -- dynamical and accretion effects. We find that the
luminosity effects, besides reprocessing of the stellar light into the
infrared region, are unlikely to be important due to existing
observational constraints. The mass effects are much more likely to be
important especially if the disk inner regions are ``light'' as is the
case when the disk is unstable to a limit cycle instability
(e.g. Siemiginowska, Czerny, \& Kostyunin 1996; Lin \& Shields 1986) or
is emptied out in the inner part for other reasons.

\section{self-gravity in quasar disks}

We shall not discuss the implications of self-gravity for quasar disks
because these have been investigated previously by a number of workers
(e.g. Paczy\'nski 1978; Kolykhalov \& Sunyaev 1980; Shlosman \&
Begelman 1989; Goodman 2003). Quasars are believed to be accreting
near the Eddington limit and hence their accretion disks need to be
quite massive.  The disks are expected to form stars at radii for
which the Toomre (1964) parameter satisfies
\begin{equation}
Q = \frac{c_s \Omega}{\pi G \Sigma}\simeq \frac{\Omega^2 H}{\pi G
\Sigma} = \;\frac{H \mbh }{\pi \Sigma R^3} \approx\;
\frac{\mbh}{\md(R)}\,\frac{H}{R} < 1
\end{equation}
where $c_s$ is the isothermal sound speed of the disk, $\Omega$ is the
Keplerian circular rotation frequency, $\Sigma(R)$ is the disk column
depth at radius $R$, $H$ is the disk half-thickness (pressure
height-scale), and finally, $\md(R) = \Sigma \pi R^2$ is approximately
the disk mass within radius $R$. Following Goodman (2003), the radii
which satisfy this conditions in the standard accretion disk are
\begin{equation}
R > R_{Q=1}\simeq 2.7 \times 10^3 R_S \left(\frac{\alpha_{-1} \mdot^2
}{M_8}\right)^{2/9}\;,
\label{rq1}
\end{equation}
where $\alpha_{-1}$ is the normalized Shakura-Sunyaev viscosity
parameter, $\mdot = \dot{M}/\medd$, where $\medd = \ledd/\epsilon
c^2$, with radiative efficiency $\epsilon=0.1$.  In this paper we
shall use the notations in which $\alpha_{-1}\equiv \alpha/10^{-1}$,
$\mdot_{-3} \equiv \mdot/10^{-3}$, etc.

It has been argued that, in addition to $Q \simlt 1$, a second
condition for star formation by self-gravity in accretion disks must
be fulfilled: the cooling time should be shorter than a fraction of
the Keplerian rotation period (e.g. Shlosman \& Begelman 1989; Gammie
2001). However, recent simulations by Johnson \& Gammie (2003)
demonstrated that the instability domain is much larger. In the
non-linear instability stage, the local cooling time may be orders of
magnitude smaller from that found in the unperturbed disk
model. Therefore Johnson \& Gammie (2003) concluded that it is
extremely difficult to prevent even the marginally stable, $Q \sim 1$,
optically thick disk from fragmenting and forming stars. Since the
formation of gravitationally bound structures is extremely fast
(Toomre 1964), it is then feared that most of the disk mass would be
used up to build up stars rather than feed the SMBH. Sirko \& Goodman
(2003) demonstrated that the amount of heat required to stop star
formation in the disk outer regions is excessive (i.e. larger than
$\ledd$ for a typical quasar), and moreover the observed quasar SEDs
rule out such a heating. Thus it seems inevitable that stars will form
in quasar disks at radii $\simgt 10^3 R_S$. Having formed there, the
stars can migrate (in-spiral) radially through exchange of the angular
momentum with the disk (e.g. Syer, Clarke \& Rees 1991), so their
radial distribution may evolve significantly on time scales that are
long enough.

It is not clear how much mass would be converted to stars in a single
accretion outburst episode, but if we assume that this is a fraction
$q < 1$ of the disk mass, then obviously
\begin{equation}
M_* N_* = q \md\, \approx \frac{q}{Q} \,\frac{H}{R}\mbh \simeq 3\times
10^5 \msun \, \frac{q}{Q} \, M_8\;,
\end{equation}
where we set $H/R \sim 3 \times 10^{-3}$ (see equation \ref{hr}),
$N_*$ and $M_*$ are the total number of stars created and a typical
stellar mass, respectively. It is also unclear how exactly this
stellar distribution evolves with time under the interaction with the
accretion flow, since stars may be swallowed or tidally disrupted by
the SMBH if they migrate close enough (Syer et al. 1991), they may
explode as supernovae (e.g. Syer et al. 1991; Rozyczka, Bodenheimer,
\& Lin 1995), destroyed by direct star-star collisions, and evolve
under the influence of long-range two-body interactions (e.g. Bahcall
\& Wolf 1976).

It is outside the scope of this paper to discuss ways for quasars to
maintain high enough mass transfer rates into the SMBHs when the
self-gravity sets in their accretion disks. Now, suppose that the
quasar mass supply has substantially decreased, so that most of the
gaseous disk mass has been used up, and that some number of the
``original'' self-gravity created stars is still caught up in the
disk. The quasar presumably becomes a LLAGN. What would then be the
role of such stars?

\section{direct contribution to LLAGN spectrum}\label{sec:rad}


Even with the HST, except for very few nearby galaxies, galactic
centers are not resolvable to scales better than $\sim 1$ pc in the
optical/UV. Stars formed in a self-gravitating disk at a distance of
$R = 0.04 (\mbh/4\times 10^6 \msun) r_5$ pc, where $r=R/R_S$, and $r_5
= r/10^5$, cannot be directly separated out from the accretion disk
spectrum. In addition, the galactic centers are viewed through a large
column depth of galactic stars (i.e. the galactic bulges). However,
Filippenko \& Sargent (1985), and Ho, Filippenko \& Sargent (1997),
created a large template of galaxies without or with very weak
emission lines. Using a sophisticated technique to fit the absorption
lines in the spectra of LLAGN, they were able to subtract out
contribution of galactic stars, and to obtain the nuclear (very weak
compared to the galaxy) continuum and broad emission lines of LLAGN.

\begin{figure}
\centerline{\psfig{file=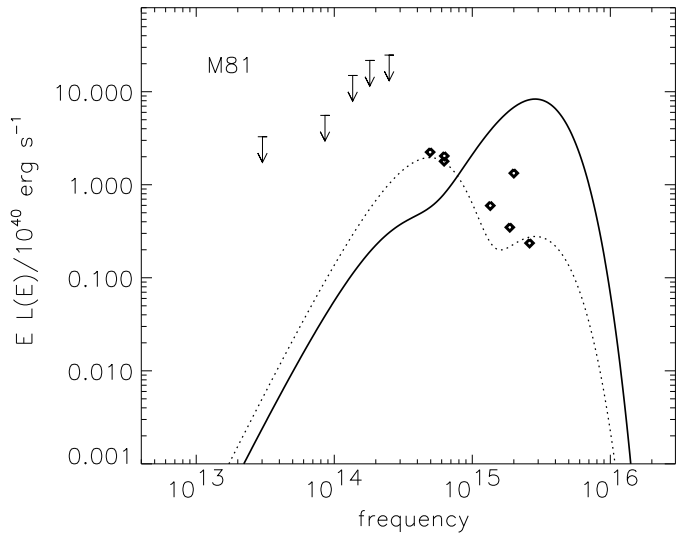,width=.5\textwidth,angle=0}}
\caption{M81 infrared to UV SED compared to the spectrum of the
central parsec of the Milky Way (solid curve). The dotted line shows
the spectrum of a nuclear star cluster that ``fits'' the M81
data. Note that the latter model requires too high a spatial density
of stars and hence it is ruled out on physical grounds (see text).}
\label{fig:mw}
\end{figure}

The stars that we are interested in are very close to the SMBH, and
their circular Kepler velocities are $v_K(r) \approx 2000\, r_4^{-1/2}
\approx 2000\, (0.1\,\hbox{pc}\, M_8/R)^{1/2}$ km/sec, where $r_4 =
r/10^4$. Since the galaxy template contains spectra with FWHM
velocities of $\simlt 500$ km/sec (Ho et al. 1997), it is most likely
that the Doppler-broadened spectra of the inner stars in the LLAGN
would be counted as the continuum emission (Ho 2004, private
communication). Can such stars make a non-negligible contribution to
the LLAGN nuclear spectrum? Two cases exist.

\subsection{Hot massive young stars}

The stellar spectrum of the inner central parsec of our GC is
dominated by hot young stars. It is thus interesting to compare its
spectrum with the better of the available nuclear LLAGN spectra.  M81
is a nearby LLAGN, with the SMBH mass estimated at a value very close
to that of \sgra (e.g. \schodel et al. 2003; Ghez et al. 2003),
i.e. $\mbh = 4\times 10^6 \msun$. M81 nuclear infrared to UV spectrum
is shown as thick data points and upper limits in Figure \ref{fig:mw}
as compiled in table 2 of Ho (1999). Ho (1999) argues against a
significant intrinsic dust extinction in LLAGN. The spectrum of stars
in the inner 1 pc in our GC is relatively well known (Mezger et
al. 1996): there is a contribution of ubiquitous low mass stars with
effective temperature $\teff = 4 \times 10^3$ K, with a combined
luminosity of $10^6 \lsun$, and that of $\sim 30$ hot young stars with
the total luminosity of $3\times 10^7 \lsun$ and the effective
temperature of $\teff=3.5 \times 10^4$ K. This spectrum is shown as a
solid curve in Figure \ref{fig:mw} and assumes no extinction (of
course we actually observe the GC through some $\sim 30$ magnitudes of
extinction in the visual).

Clearly, had there been a similar amount of such hot luminous young
stars in M81 as in the GC, these stars would easily out-shine the LLAGN
in M81 in the UV band. However in the GC the vast majority of the hot
young stars move with velocities of order $\simlt 500$ km/sec, so that
the subtraction technique of Ho et al. (1997) would have probably
removed them from the nuclear spectrum. In any event, stars moving
with so ``low'' velocities are outside the radius where most of the
accretion flow physics is played out ($\simlt 10^5 R_S$). The results
of Ho (1999) thus clearly show that there are no or very few hot young
luminous O-stars in the inner accretion flow region in most of the
LLAGN (LINERS) as otherwise they would produce a component similar to
the BBB.

Since the history of star formation and evolution in an accretion disk
is extremely model- and parameter-dependent, it is possible that such
young massive stars never formed in the LLAGN accretion disk. It is
also possible that these stars have already gone through the supernova
phase (e.g. Syer et al. 1991), or are on the red giant branch, thus
shifting their spectrum from the UV into the optical frequencies (but
see the next section).

\del{Below we also show that the high mass stars embedded in the accretion
disk could be obscured from our direct view by the bloated disk if
observed at significant inclination angles to the nuclear disk (which
may be different from the galaxy plane). In addition, for \sgra,
Genzel et al. (2003) argued that the mid-infrared excess in the
spectrum of a close $\sim$ O-type star (named S2) may be explainable
by the presence of dust embedded in the hot (e.g. non-radiative)
accretion flow at distances $\ll 1$ pc from \sgra. For higher hot gas
densities the dust will be more efficiently destroyed by sputtering;
nevertheless, this effect deserves a more careful parameter range
study.}

\subsection{Low mass stars or evolved red giant stars}

The dotted line in Figure \ref{fig:mw} is a very simple model that
``fits'' the M81 data much better. In this model the ``low'' mass
stars have average temperature of $6 \times 10^3 $ K, with the total
luminosity of $6 \times 10^6\lsun$, and just one hot luminous star
with $L=10^6 \lsun$ and $\teff =3.5 \times 10^4$ K. Clearly, a more
realistic model with a power-law mass function could be designed that
would easily fit the ifra-red to UV SED of M81 with no contribution
from the accretion onto the SMBH whatsoever.

However, there are several arguments against this possibility. As we
argued above, to appear as AGN continuum, the spectrum of such stars
would have to be strongly smeared, say with random velocities of order
$2000$ km/sec or more. For M81, the mass of the stars would exceed the
deduced SMBH mass, thus leading to a direct contradiction with
observations. Further, the stellar densities in the inner part of the
stellar cluster would be as large as $10^{14} \msun \hbox{pc}^{-3}$ if
the stars are main sequence stars with $M\sim \msun$. Such high
stellar densities are extremely unlikely because stellar physical
collisions and disintegrations would be expected (e.g. Murthy, Cohn,
\& Durisen 1991).  If the stars are much more luminous super-giants,
then the required stellar number density is much lower but because of
their large radii, we find that the collision time is of the same
order, i.e. too short. The red giant stars are also easy targets for
ablation of mass due to star-disk collisions (Armitage , Zurek \&
Davies 1996). Finally, there is a deficit of such stars in the inner
cusp of our GC (Genzel et al. 2003).

\subsection{Infrared spectrum: reprocessing in the disk?}

The stellar cusp/cluster could also contribute to heating the gaseous
disk, especially its surface layers. If the stars are uniformly
distributed within radius $R$, then the effective temperature of the
stellar radiation is $63 L_{41}^{1/4} (R/\hbox{pc})^{-1/2}$ Kelvin,
where $L_{41}$ is the total stellar luminosity in units of $10^{41}$
erg sec$^{-1}$. We argued that it is not plausible for the stars to
dominate the nuclear optical/UV continuum of LLAGN, hence it seems
unlikely that they would significantly add to the reprocessed disk
emission.

This may not be the case, however, because the accretion disk is very
thin (equation \ref{hr}), and only a very small fraction of the
luminosity generated by accretion onto the SMBH may impact the outer
disk. In contrast, the accretion disk is geometrically thick compared
to stars and would reprocess a large fraction of stellar light even if
stars are exactly in the disk midplane. If stars are on more general
orbits (see \S \ref{sec:hi}), then roughly half of their radiation is
reprocessed in the disk, possibly making it important in the
mid-infrared region where a ``bare'' disk (without surrounding stars)
is predicted to be dim. In this case the distribution of the radiation
impinging on the disk will be quite patchy. Therefore there may be a
broad range of temperatures resulting from the stellar illumination of
the disk which would obviously lead to a broad infra-red feature in
the SED of the LLAGN. This feature will be spectroscopically much
broader than that resulting from the disk illumination by the single
central source (the accreting SMBH), and it may be also variable on
time scales of weeks to years because of bright stars approaching and
passing through the disk (e.g. Nayakshin \& Sunyaev 2003; Cuadra et
al. 2003). Stars also heat the disk via bodily impacts (e.g. Zentsova
2003, Norman \& Silk 1983) and can increase the effective disk
viscosity (Ostriker 1983).

\section{Embedded stars and disk accretion}\label{sec:accretion}

Stars orbiting the SMBH in the inner parsec of the accretion flow will
also modify the accretion flow of gas. We assume that the combined
mass of stars in this region is much smaller than $\mbh$, and consider
the case when the stellar and disk planes coincide exactly. In this
limit the relative velocity of a star and the gas in the disk
surrounding the star is negligible. Thus star-disk interactions are
qualitatively different. The interaction is best understood based on
the analogy between this case and that of a better studied
circumstellar disk with embedded (proto-) planets. If the star does
not open up a radial gap in the disk (see below), then it grows via
Bondi-like gas accretion. If the gap is opened, the accretion onto the
star slows down significantly. The star however continues to interact
with the disk due to gravity (essentially via dynamical friction). In
both gap/no-gap cases the star can migrate, that is spiral-in,
radially. Clearly the flows of gas and stars are not independent.

\subsection{Gap formation conditions}

A star co-rotating with the disk interacts with the gas in the disk
via gravitational torques (e.g. Goldreich \& Tremaine 1980; Lin \&
Papaloizou 1986a,b). The linear velocity of gas at radii greater than
the radius at which the star circles the SMBH, $R$, is increased as a
result of the interaction with the star. Thus its angular momentum
increases and it moves outward from the star's location. The opposite
happens to the gas interior to the star's location. Hence the star
pushes the gas away from its radial position. If this effect is strong
enough, a radial gap is opened in the disk. 

The Hill's radius $R_H$,
\begin{equation}
R_H = \left[\frac{M_*}{3\mbh}\right]^{1/3}\,R\;,
\label{rh}
\end{equation} 
defines the star's sphere of influence in the disk: dynamics of gas
within the sphere is dominated by the star rather than the SMBH.

There are two conditions for formation of the radial gap in the
disk. The first is that $R_H$ is greater than the pressure scale
height of the disk, $H$. Otherwise the gas would be able to overflow
the star at vertical heights $z > R_H$.  The disk height-scale $H$ for
gas-pressure standard accretion disk at $r\equiv R/R_S\gg 1$ is
\begin{equation}
H/R = 2.5 \times 10^{-3} \left(\alpha M_8\right)^{-1/10} r^{1/20}
\mdot^{1/5} \;
\label{hr}
\end{equation}
(see Svensson \& Zdziarski 1994). We assume that the accretion disk is
not strongly self-gravitating during the LLAGN epoch, i.e. that
$Q\simgt 1$. Since
\begin{equation}
\frac{H}{R_H} = 0.26 \left(\alpha M_8\right)^{-1/10} r^{1/20}
\mdot_{-4}^{1/5} \left(\frac{M_8}{m_*}\right)^{1/3} \; < 1\;,
\label{hrh}
\end{equation}
where $m_*\equiv M_*/\msun$.  The first condition for a gap formation
is usually satisfied in LLAGN disks.

The second criterion is that the rate with which the gas is pushed
away from the star's location due to gravitational interaction is
higher than the rate with which the gap could be filled in due to
viscous gas transport. This is possible only when the disk viscosity
is low enough. Parameterized in terms of the Shakura-Sunyaev
$\alpha$-parameter, this condition becomes
\begin{equation}
\alpha < \alpha_{\rm min} = \frac{1}{40} \,
\left(\frac{M_*}{\mbh}\right)^2\, \left(\frac{R}{H}\right)^5
\label{amin}
\end{equation}
(e.g. Syer et al. 1991). Using the expression for $H/R$, we get
\begin{equation}
\alpha_{\rm min} = 3.8\times 10^{-5} \frac{m_*^4}{M_8^3}\, r_2^{-1/2}
\mdot_{-3}^{-2}\;.
\label{amin1}
\end{equation}
It is unlikely that $\alpha$ is as low as this, and hence the low mass
stars are unable to open a gap in the disk of a ``large'' SMBH,
i.e. $M_8\simgt 1$. However, for $\mbh = 4 \times 10^6 \msun$, the
coefficient in equation \ref{amin1} is $\alpha_{\rm min} = 0.6$: low
mass stars may open a gap in accretion disks of lighter galactic
nuclei. Finally, for $M_8=1$ and $m_*=10$, i.e. a high mass star with
$M_*=10 \msun$, we have $\alpha_{\rm min} = 0.38$, thus a gap formation
is again likely.

\subsection{No gap: accretion onto embedded stars. Forming an inner
hole?}

Consider now the cases when the stars are unable to open up a gap. As
shown above, a gap is unlikely to form for high mass black holes,
i.e. $M_8 \simgt 1$, or when the stars are low mass stars. Also, gap
formation is more problematic at large radii $r_2 \gg 1$.  The stars
then grow by accretion of gas from the disk (Syer et al. 1991).

\subsubsection{Low mass stars}

For low mass stars, stellar radiation and winds are relatively
weak. The accretion of gas onto a low mass star is then similar to the
growth of terrestrial planets in a planetesimal disk (Lissauer 1987;
Bate et al. 2003; Tan \& Goodman 2004). Since $R_H > H$, accretion
onto the star is quasi two-dimensional. The accretion rate is
determined by the rate at which differential rotation brings the
matter into the Hill's sphere,
\begin{equation}
\dot{M}_* = \dot{M}_H \sim 4 \pi R_H H\, \rho \,v_H \sim 4 \pi R_H^2
\rho c_s \;,
\label{mhill}
\end{equation}
where $\rho = \Sigma/2H$ is the mean disk density. We used the fact
that the characteristic gas velocity (relative to the star) at the
Hill's distance from the star, $v_H$, is $v_H = R_H |d \Omega/d\ln R|
\sim c_s (R_H/H)$ since the angular velocity for Keplerian rotation is
$\Omega = c_s/H$.  Equation \ref{mhill} is valid as long as $R_H > H$
since in the opposite case the gas thermal velocity becomes important
and the accretion would proceed at the Bondi rate (e.g., Syer et al
1991).  Of course $\dot{M}_*$ cannot exceed the Eddington limit value,
e.g.,
\begin{equation}
\dot{M}_{*, \rm Edd} \sim 10^{-3}\; \frac{m_*}{r_*} \msun
\; \hbox{year}^{-1}\;,
\label{med}
\end{equation}
where $r_*\equiv R_*/\rsun$, star's radius in Solar radii units.

\subsubsection{High mass stars in the disk}\label{sec:highmass}

The enormous radiation field of luminous high mass stars pre-heats the
disk material at distances $\sim R_H$ from the star. If the
temperature established is much higher than the unperturbed disk
temperature, the flow of gas will be strongly affected. Indeed,
$\teff(R_H) = [L_*/4\pi \sigma R_H^2]^{1/4}$,
\begin{equation}
\teff(R_H) = 730 \;\hbox{K}\; \frac{L_5^{1/4} M_8^{-1/3}
r_4^{-1/2}}{(M_*/30\msun)^{1/6}}\;,
\label{teffh}
\end{equation}
(where $L_5 \equiv L_*/10^5 \lsun$) can be larger than the disk effective
temperature, $T_{\rm eff, d}$:
\begin{equation}
T_{\rm eff, d} = 110 \;\hbox{K}\; \left(\frac{\mdot_{-4}}{M_8
r_4^3}\right)^{1/4} \;.
\label{teffd}
\end{equation}
The accretion disk next to a high mass star is thus bloated. At
distances from the star $\simlt R_H$, the star's gravity takes over
and a thin small scale circum-stellar disk develops. Perpendicular to
the disk, strong stellar winds sweep up the colder matter. A sort of
torus thus forms around the star. There will be a thin ionized
(Stromgren) layer on the side of the torus facing the star. At low
disk surface densities appropriate to LLAGN, high mass stars loose
rather than gain mass from the disk; the winds may cool and condense
on the disk surface\footnote{Note that neither the hot stellar wind
nor the gas evaporated from the disk escape to infinity because the
escape temperature is $\sim T_{\rm vir}\sim 3 \times 10^8 \hbox{K}
r_4^{-1}$}. The accretion rate onto the massive star is strongly
reduced due to the disk pre-heating. The disk accretion on the high
mass stars is completely negligible compared with that on the much
more numerous low mass stars that do not suffer from pre-heating and
stellar winds.

The observational appearance of the star will be strongly angle
dependent. Viewed pole-on to the disk, the intrinsic stellar emission
is clearly visible, as is the line emission resulting from the disk
illumination. However, when viewed at large inclination angles, the
stellar emission is hidden from the observer similar to the central
regions of type 2 AGN. The optical line emission of the inner part of
the ``torus'' around the star may still be visible.

Note that for quasar disks, $\mdot\sim 1$ and then $T_{\rm eff,
d}\simgt \teff(R_H)$ even for the brightest O stars. In addition, for
smaller radii, i.e. $R \sim 100 R_S$, the disk pre-heating is
unimportant even for LLAGN.

\subsubsection{Eating the disk from within}\label{sec:hole} 

As shown previously, the embedded stars gain mass from the accretion
disk. Here we consider only the low mass stars that are certain to
dominate the disk-stars mass exchange due to their large number. From
the standard accretion theory, the gas accretion rate onto the SMBH is
\begin{equation}
\dot{M}_{\rm BH} = 4 \pi R H \rho |v_R| \sim 4 \pi H^2 \rho \alpha c_s
\end{equation}
since $|v_R|\sim \alpha (H/R) c_s$ (e.g. Frank, King \& Raine
1992). Therefore the rate of accretion on the single star compares to
the accretion rate onto the black hole as
\begin{equation}
\frac{\dot{M}_H}{\dot{M}_{\rm BH}} \simeq \frac{R_H^2}{\alpha
H^2}\simgt 1\;\; \hbox{!!}
\end{equation}
Surprisingly, just one star in a LLAGN disk may consume the disk
quicker than the SMBH, leaving little for the SMBH growth. Of course
the accretion onto the proto-star is limited to $\dot{M}_{*, \rm
Edd}$, for one. Pre-heating of the gas by the radiation from the star
and the circum-stellar accretion disk may decrease the accretion rate
further. Nevertheless, even if accretion on a single star is very
inefficient compared with equation \ref{mhill}, a large number of
proto-stars in the disk can make up for this, and thus the combined
mass accretion on these stars could well exceed that onto the SMBH
during the LLAGN phase.

\begin{figure}
\centerline{\psfig{file=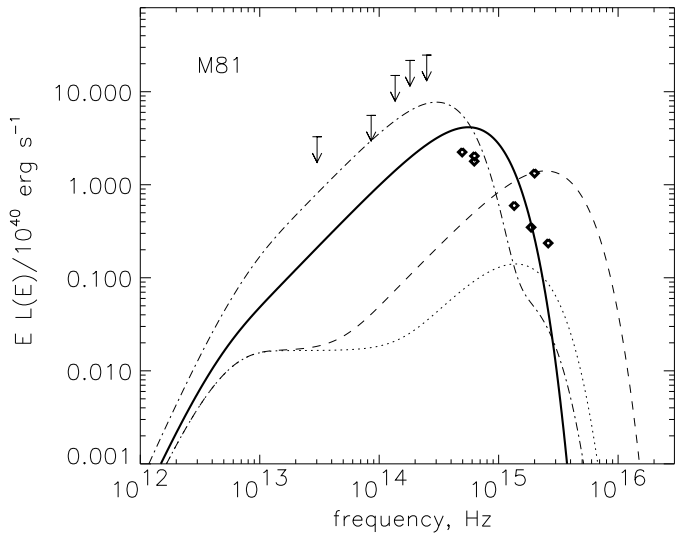,width=.5\textwidth,angle=0}}
\caption{Modified cold disk fits to M81 infrared to UV SED. The
standard disk with the inner cutoff radius of $100 R_S$ is shown with
the thick solid curve (Quataert et al. 1999 model). The rest of the
curves assume cold inner disk persisting to the last stable orbit but
the accretion rate decreasing with radius as prescribed (see text for
detail).}
\label{fig:gradual}
\end{figure}

\subsection{Carving out an inner hole?}

If stars embedded inside the disk eat it away before the gas can reach
the SMBH, then the disk inner region may in principle be almost empty
and hence underluminous compared with the steady-state standard
accretion disk. This could then explain the observed lack of the BBB
in the LLAGN spectra, interpreted as an inner hole in the accretion
disk (e.g. Quataert et al. 1999).

We test this idea by comparing the expected disk spectra to the
observations of M81 as an example.  Given the quality of the data, we
model the SED of cold disks consumed by the stars with a very simple
model. We use the standard multi-temperature black-body spectrum with
the exception that the accretion rate, $\dot{M}(R)$, is a function of
radius. We neglect stellar irradiation compared with internal disk
viscous heating. The local disk effective temperature is then given by
equation \ref{teffd}.

We assume that stars are only present at radii larger than a
``self-gravity'' radius, $\rsg$. The latter may be smaller than
$R_{Q=1}$ because of the possible radial migration of stars. For
simplicity, we choose $\mdot(R) = \mdot(\rout) (R/\rout)^p$, for $\rsg
< R < \rout$, and $\mdot(R) = \mdot(\rsg)$ for $R < \rsg$. Figure
\ref{fig:gradual} shows the same data as Figure \ref{fig:mw} with
several models for the disk emission. The solid line shows the
alternative cut-off disk model similar to Quataert et al. (1999) --
the standard accretion disk with the inner radius of $R_i = 100 R_S$
and the disk accretion rate $\mdot(\rout) = 4 \times 10^{-3}$. The
cold flow is presumed to evaporate and not to emit in the optical/UV
region for $R < R_i$. The dotted line shows the model with $p=1$,
$\rsg = 100 R_S$, $\rout = 10^5 R_S$ and $\mdot = 4 \times
10^{-3}$. Since for $p=1$ the energy emitted per a decade in radius
($\propto G\mbh\dot{M}(R)/R^3 \times \pi R^2$) becomes independent of
radius, there is a flat part in the infrared frequency part ($\nu \sim
10^{13}-10^{14}$ Hz) of the spectrum for this model. The model is too
dim in the optical/UV. The dashed line shows a similar model but with
$\rsg = 10^3 R_S$. The model is clearly incompatible with the data.

Finally, the dot-dashed curve shows a more complex model with $\mdot
=$~const~$=3\times 10^{-2}$ for $500 R_S = \rsg < R < \rout$, $\mdot =
\mdot(\rsg) (R/\rsg)^6$ for $ 100 R_S < R < \rsg$, and $\mdot(R) =
\mdot(\rsg)$ for $R < \rsg$. $\rout=10^5 R_S$ in the model. This model
roughly ``fits'' the data. However, is is clearly fine-tuned to have
a similar inner radius of $100 R_S$ as the model of Quataert et
al. (1999).  The ``eaten-from-within'' disk model may fit the data
only if the accretion rate at the region $R\sim$~few hundred~$R_S$ is
similar to the cut-off disk model, and then drops very sharply. (What
happens at larger radii is not that important as long as the outer
disk is not too luminous in the infra-red region). A sudden disk
depletion at a small range of radii is not impossible if there is a
large concentration of low mass stars there, but is seems a rather
unlikely case to us. The total mass of the stars in this disk
depletion region would not be very large, $\delta M_* \sim
\dot{M}(\rout) t \simeq 10^3 \msun (t/10^7$~years) for $\mdot = 4
\times 10^{-3}$. However, the radial size of the $100 R_S$ region is
only $\sim 10^{14}\, \hbox{cm} = 3 \times 10^{-5} pc$, for the M81
SMBH mass, yielding stellar density of order $3 \times 10^{16}
\msun/\hbox{pc}^3$. Any star present at such small radii from a
pre-existent isotropic cluster would have been destroyed in only $\sim
3$ years. The resulting disintegration of the stars would have
resulted in a very energetic accretion event. Thus this picture would
mandate periods of quiescence when stars steal most of the mass from
the accreting flow, and shorter periods of strong accretion activity
onto the SMBH when the stellar collisions or supernova explosions (if
stars grow to be too massive) release the stored gas. As time between
these luminosity variations is long compared to history of Astronomy,
it is not immediately clear whether such model is supported by
observations or not.

\subsection{Gap accretion. Clogging accretion flow
with embedded stars?}\label{sec:load}

For higher stellar mass to the SMBH mass ratios, gap formation occurs
(see eqs. \ref{amin} \& \ref{amin1}). The accretion rate of the star
is reduced by the gap presence, but does not stop completely
(e.g. Bate et al. 2003). When the mass of the star is small compared
to the surrounding disk, the radial motion of the star is similar to
that of the fluid at the same position in the disk. The star thus
migrates closer to the SMBH on the disk viscous time scale. Because of
its small mass, the star does not significantly alters the radial
surface density profile of the disk (Lin \& Papaloizou 1986b). While
the detailed disk spectrum may be affected by the presence of the
star, the effect is quite minor as long as the star's mass is small
compared to that of the disk, and the disk is optically thick.

However, when the mass of the star (or any massive satellite such as a
secondary black hole) is considerable, the accretion disk is bound to
``notice'' the satellite (Syer \& Clarke 1995). Since the angular
momentum exchange continues through the gap, the disk interior to the
star is able to continue accretion onto the SMBH. At the same time,
the gas at radii larger than the position of the star is blocked by
the gap. Because of the high inertia of the star, the radial velocity
of the gas just outside of the gap is smaller than its unperturbed
(i.e. star-free) value, and the disk surface density builds up there
(Lin \& Papaloizou 1986b; Syer et al. 1991; Ivanov, Papaloizou, \&
Polnarev 1999). Syer \& Clarke (1995) show that the radial migration
velocity of the satellite becomes
\begin{equation}
v_s = v_{R, d} \frac{2 \pi \Sigma R^2}{M_*}\;, 
\end{equation}
where $v_{R, d} = 3 \alpha (H/R)^2 v_K(R)$ is the disk unperturbed
radial viscous velocity. If $M_* \gg \pi \Sigma R^2$, when the disk
accretion is basically temporarily stopped, until the mass behind the
satellite becomes large enough to push it through into the black
hole. Meanwhile, the inner disk, cutoff from the mass supply, empties
out and may become quite dim (Syer \& Clarke 1995). This situation
will persist during time $t \sim M_*/\dot{M} \sim 10^3 \, m_* M_8^{-1}
\mdot_{-3}^{-1}$ years. This is actually an upper limit, because the
overflow of the gap may occur at lower values of $\Sigma$, i.e. at
somewhat earlier times. After enough mass has been accumulated, the
accretion resumes, resulting in a bright outburst. The star and the
built-up disk accrete onto the SMBH on the disk viscous time scale,
the one that reflects the amount of the stored mass. If the accretion
rate in such an outburst is high enough, the inner disk becomes
radiation-pressure dominated and the viscous time may be significantly
shorter than that of the quiescent disk at the same location (this is
due to a change in the factor $(H/R)^2$). The activity phase is thus
relatively short compared with the mass accumulation phase.

Are such blocked accretion flows relevant to the apparently cutoff
disks of LLAGN? If this were the case, the LLAGN activity would be
periodic, with the LLAGN being the quiescent state and a more brighter
phase, such as Seyfert galaxies nuclei, would be the outburst
state. The mass of the unperturbed accretion disk, estimated as $\md
\sim \pi R^2 \Sigma(R)$ in the standard accretion flow is
\begin{equation}
\md = 60 \msun \alpha^{-4/5} M_8^{11/5} r_2^{7/5} \mdot_{-3}^{3/5}
\;. 
\label{mdisk}
\end{equation}
The disk is too heavy for a single star to induce serious accretion
rate modulation for these parameters. For a smaller SMBH mass, say
$4\times 10^6 \msun$, the coefficient in equation \ref{mdisk} is about
$0.05 \msun$, making a single star a big drag onto the accretion
flow. The accretion luminosity can then be substantially
modulated. Syer et al. (1991), considering the case of a $10^6 \msun$
SMBH, already mentioned such a possibility. The quiescence time can be
as long as $10^5$ years in this case.

Therefore, the possibility of a significant time variability due to
presence of stars in the disk, and the ``empty'' observational
appearance of the inner disk, cannot be ruled out for lower mass
black holes. However, the absence of the UV bump (BBB) in the LLAGN
spectrum is quite a general feature, seemingly independent of the SMBH
mass. Indeed, many of the objects in the Ho (1999) sample have SMBH
masses higher than $10^8\msun$. The recent observations of Ptak et
al. (2004) of NGC 3998, a LLAGN with the SMBH estimated at $\sim 10^9
\msun$, also shows no clear BBB. Many stars embedded in the accretion
flow and all drifting together could slow down the accretion in the
inner regions of high SMBH accretion disks, but one would naively
expect the stars to be smoothly distributed in the disk.  Hence
embedded stars cannot explain cutoff disks in high mass
LLAGN. Therefore clogging up the disk flow with embedded stars is
possible, but it cannot be the reason behind the missing inner
accretion disk of LLAGN.

\section{Stars at high inclination angles to the disk}\label{sec:hi}

So far we have been concerned with stars embedded in the disk. In
general, however, there will be stars on highly inclined (to the disk)
orbits, brought in by relaxation of the old central star cluster
(e.g. Bahcall \& Wolf 1976), or from partial capture of stars by the
accretion disk (Syer et al. 1991), etc. If stellar orbital planes and
the disk plane are significantly mis-aligned, then most of the
interactions take place when the stars hit the disk and pass through
it (e.g. Ostriker 1983; Norman \& Silk 1983; Zentsova 1983; Perry \&
Williams 1993; Nayakshin et al. 2004). The stars move through the disk
at very high Mach numbers and produce strong shocks. Quite generally,
the stars are expected to enhance (speed-up) the accretion flow
because of the additional angular momentum transfer mechanism
(Ostriker 1983) and the extra heating. Thus this effect works in the
opposite direction to that discussed in \S \ref{sec:load}.

For radii smaller than some value $R_i$, the angular momentum transfer
in the disk is dominated by star-disk collisions rather than by the
internal disk viscosity. The value of $R_i$ is a function of the
number density of stars in the inner part of the stellar cusp and
their orbits (see Nayakshin et al. 2004 for estimates for \sgra\ under
assumption of an isotropic stellar cusp). If the angular momentum
transfer mediated by stars is very fast, then the surface column depth
of the disk is much smaller (for the same $\dot{M} = 2 \pi R \Sigma
v_R =$ const) than it is for the standard disk. If it becomes
optically thin, the disk will be much hotter for the same accretion
rate. In fact, if it is unable to cool, the disk switches to a hot
accretion mode, opening up an inner hole in the optically thick
geometrically thin disk\footnote{Interestingly, this remark is also
relevant to accretion disks around stellar-mass compact objects
interacting with a cloud of comets -- see Pineault \& Landry 1994 and
references therein}.

Due to interactions with the disk, embedded stars have Keplerian
circular velocities. If two such stars meet in the disk, their
relative velocity is small. In contrast stars out of the plane are
expected to have a broad velocity distribution (e.g. Bahcall \& Wolf
1976) and are prone to self-destruction in hard collisions (e.g. see
\S 3 of Murthy et al. 1991, and references therein). Such a collision
may involve two highly inclined stars or one star embedded in the disk
and the other at a highly inclined orbit.  If such an event happens in
the inner disk region, and especially if the mass of the inner disk is
not too much greater than the combined mass of the two stars, then the
resulting energy release will be sufficient to significantly speed up
the accretion of the inner disk. The inner disk may then be quickly
emptied out, and while it is being refilled the disk will radiate as a
cut-off disk.

As did the effects discussed in \S \ref{sec:load}, these two
mechanisms for the inner disk removal are also of a temporary
nature. Unless being flat as an accretion disk, the inner stellar cusp
will self-destruct by binary collisions in a relatively short time,
e.g. $10^7-10^8$ years for typical cluster parameters (see, e.g.,
Fig. 6 of Murthy et al. 1991). Hence these mechanisms for the inner
disk removal can be relevant to LLAGN only if these are relatively
short lived phenomena in the SMBH/AGN activity cycles.

\section{Discussion and Conclusions}

Here we suggested that close stars ($R\ll 1$ pc from the SMBH) created
in the accretion disk by self-gravity in the bright quasar phase, may
be present in the LLAGN disks. We found (\S \ref{sec:rad}) that it is
unlikely that these disk stars significantly contribute to the LLAGN
optical/UV spectrum directly.  Hot young massive stars are ruled out
because most LLAGN are weak in the UV; and low mass stars would have
to be too numerous. Stellar contribution in the infra-red range is
plausible, however, because SED of standard accretion disks are weak
in that region.

The effects of stars' mass, i.e. inertia, are much more realistic
because they require a ``small'' number of low-mass stars embedded in
the disk (say $10^2-10^4$ stars). These stars can consume the disk
faster than it can accrete on the SMBH (if no radial gap forms in the
disk). In this case little gas may be flowing into the central
SMBH. However, when fitting the optical/UV spectrum of a ``typical''
LLAGN, M81, we found that this model would only work if the disk were
quickly and abruptly consumed by the embedded stars in the inner
$\sim$ few hundreds \sch radii. Further work is needed to determine
whether this is natural and may be sustained for long enough times to
explain the LLAGN phenomenon.

If a radial gap does form, then stars embedded in the disk do not
accrete gas from the disk. However their presence will be still felt
by the disk if the disk mass interior to the star is smaller than the
mass of the star (or group of stars). This appears to be especially
important in the quiescent phase of limit-cycle unstable disks, in
which case the stars will clog-up the inner accretion flow for time
necessary for accumulation of enough disk matter to push the star in
the disk. Since the disk interior to the star's location is quickly
emptied out (Syer \& Clarke 1995), such a star-terminated disk would
appear as a cut-off disk.  Finally, we noted that hard stellar
collisions in the inner disk region may disrupt the disk -- create
temporary inner holes.

We shall however notice that it is unlikely that the star-disk
interactions discussed in this paper are the only explanation to the
LLAGN ``inner holes'' in the accretion disk. As we have seen in \S
\ref{sec:load}, the character of star-disk interactions depends
strongly on the mass of the SMBH. Observations span a very large range
in the SMBH mass, i.e. almost $10^3$ (see e.g. Ho 1999, Ptak et
al. 2004). Moreover, the accreting (stellar mass) binary systems may
have inner holes in their accretion disks in the low luminosity states
as well (e.g. see Esin, McClintock \& Narayan 1997; McClintock et
al. 2003). Therefore the star-disk interactions may be an {\em
additional reason} for a complicated time-variability in a disk that
is already unstable (e.g. Lin \& Shields 1986; Siemiginowska et
al. 1996) or time-dependent because of a changing supply of fuel at
large distances, but they are unlikely to be the primary reason for
such variations in the disk structure.

Nevertheless with observations becoming progressively better, we
should expect many surprises, especially exotic variability, from the
effects of star-disk interactions.

\section{ACKNOWLEDGMENTS}

The author acknowledges very useful discussions with Jorge Cuadra,
Luis Ho, Jeremy Goodman and Marat Gilfanov.

\end{document}